\documentclass[12pt]{article}
\usepackage{amssymb}
\usepackage{epsfig}
\usepackage{graphicx}
\usepackage{amsmath}
\usepackage{verbatim}

\begin{document}
\begin{titlepage}
\title{\bf\Large Oblique Corrections in the MSSM at One Loop. II. Fermions  \vspace{18pt}}
\author{\normalsize Yao~Yu$^{1}$, Sibo~Zheng$^{1,2}$ \vspace{12pt}\\
{\it\small $^{1}$ Department of Physics, Chongqing University, Chongqing 401331, P. R. China}\\
{\it\small $^{2}$ Kavli Institute for Theoretical Physics China }\\
{\it \small Chinese Academy of Sciences, Beijing 100190, P. R. China }
}

\date{}
\maketitle \voffset -.3in \vskip 1.cm \centerline{\bf Abstract}
\vskip .3cm  This paper is the completion of an earlier work arXiv:1207.4867
which involves the derivation of oblique corrections in the MSSM at one-loop.
In terms of the two-component spinor formalism,
which is new in compared with those used in the literature,
the contributions arising from the fermion superpartners i.e, neutralino-chargino sector
to self-energy of standard model electroweak gauge bosons are calculated.
Corresponding descendants the $S$, $T$ and $U$ parameters are presented.
The validity of our results is examined in two ways,
which are then followed by detailed analysis on the results in the literature.

\vskip 5.cm \noindent July 2012
 \thispagestyle{empty}
\end{titlepage}

\section{Introduction}
As more data collected at the LHC,
more hints imply the absence of natural supersymmetry (SUSY) as the TeV-scale new physics beyond standard model (SM).
It seems possible to establish or rule out the minimal supersymmetric model by combining the present data at the LHC and other colliders.

One way to explore this issue is by analyzing the oblique corrections \cite{Peskin1, Peskin2} to electroweak observables arsing from the supersymmetric particles.
The logic is that these new states contribute to the precise electroweak observables such as the weak mixing angle $s^{2}_{W}$,
whose values depend on these new states' masses.
What is interesting is that the sensitivity to these masses (including the SM-like Higgs mass) in MSSM is quite unlike to the situation in SM \cite{YZK},
where the dependence on $m_h$ is logarithmic.
On the other hand,
more robust bounds on  superpartners masses appear at the LHC in comparison with the other high energy colliders.
To date the uncertainties for these observables can more severely constrain the allowed region for these superpartners masses
than what we have expected before.

In this paper, we complete our calculations of the oblique corrections in MSSM based on our previous work \cite{YZK}.
We follow the two-component spinor formalism \cite{Martin08}
to calculate the self-energy diagrams of vector bosons with neutralino and chargino-fermions loop.
This formalism is very uself since the charginos $\chi^{\pm}_{i}$ are Dirac,
 while the neutralinos $\chi^{0}_{j}$ are Majorana fermions.
Although more graphs need to be considered compared with the four-component spinor formalism,
it is quite straightforward to evaluate these graphs by incorporating one-loop integral functions \cite{functions}.

The paper is organized as follows.
In section 2, we briefly review the Lagrangian for the neutralino-chargino  (NC) sector.
we emphasize the notation and conventions when necessary.
In section 3, we derive the contributions in the NC sector.
To examine the results presented in section 3,
section 4 is devoted to a preliminary check via the decoupling limit.
In section 5, we derive the $S$, $T$ , $U$ parameters relevant to the corrections to precise electroweak observables.
The property of finiteness for these parameters can serve as another examination on the validity of the results.
Finally, we compare our results and those proposed in the literature,
 and make a few comments and conclusions.
An appendix is added to explicitly show the relevant Feynman rules in the NC sector for our calculations.

\section{Lagrangian For the NC Sector}
As completion we begin with a brief review on the Lagrangian for NC sector,
address the notations and conventions when necessary.
The Lagrangian for NC sector under gauge eigenstates is given by,
 \begin{eqnarray}{\label{E1}}
\mathcal{L}&=& -i \tilde{W}^{\dag a}\bar{\sigma}^{\mu}(\delta^{ac}\overrightarrow{\partial}_{\mu}+g\epsilon^{abc}W^{b}_{\mu})\tilde{W}^{c}\nonumber\\
&-&i ((\tilde{H}^{+}_{\mu})^{\dag},(\tilde{H}^{0}_{\mu})^{\dag})\bar{\sigma}^{\mu}(\overrightarrow{\partial}_{\mu}-ig'B_{\mu}-igY_{1}W^{a}_{\mu}\tau^{a})\left(\begin{array}{c}
                   \tilde{H}^{+}_{\mu}\\
                  \tilde{H}^{0}_{\mu}
                  \end{array}\right)\\
&-&i ((\tilde{H}^{0}_{d})^{\dag},(\tilde{H}^{-}_{d})^{\dag})\bar{\sigma}^{\mu}(\overrightarrow{\partial}_{\mu}-ig'B_{\mu}-igY_{2}W^{a}_{\mu}\tau^{a})\left(\begin{array}{c}
                   \tilde{H}^{0}_{d}\\
                  \tilde{H}^{-}_{d}
                  \end{array}\right)\nonumber
\end{eqnarray}
Where $W^a$ represent the $SU(2)_L$ gauge symmetry, $Y_1$ and $Y_2$ label the $U(1)_Y$ charges for the Higgs doublets.
Reorganize the freedoms in \eqref{E1} and adopt the convention for charginos and neutralinos,
 \begin{eqnarray}{\label{E2}}
\left(\begin{array}{c}
                   \tilde{C}^{+}_{1}\\
                  \tilde{C}^{+}_{2}
                  \end{array}\right)=V\left(\begin{array}{c}
                   \tilde{W}^{+}_{1}\\
                  \tilde{H}^{+}_{\mu}
                  \end{array}\right),~~~~~~~~
\left(\begin{array}{c}
                   \tilde{C}^{-}_{1}\\
                  \tilde{C}^{-}_{2}
                  \end{array}\right)=U\left(\begin{array}{c}
                   \tilde{W}^{-}_{1}\\
                  \tilde{H}^{-}_{d}
                  \end{array}\right)
\end{eqnarray}
as well as  $\psi^{T}=(\tilde{B}^{0}, \tilde{W}^{0},\tilde{H}^{0}_{d}, \tilde{H}^{0}_{\mu})$,
\eqref{E1} can be rewritten as,
 \begin{eqnarray}{\label{E3}}
\mathcal{L}&=&\left[-e\delta_{ij}(\tilde{C}^{+}_{i})^{\dag}\bar{\sigma}^{\mu}\tilde{C}^{+}_{j}+e\delta_{ij}(\tilde{C}^{-}_{i})^{\dag}\bar{\sigma}^{\mu}\tilde{C}^{-}_{j}\right]A_{\mu}\nonumber\\
&+&\frac{g}{c}\left[O'^{L}_{ij}(\tilde{C}^{+}_{i})^{\dag}\bar{\sigma}^{\mu}\tilde{C}^{+}_{j}-
O'^{R}_{ij}(\tilde{C}^{-}_{i})^{\dag}\bar{\sigma}^{\mu}\tilde{C}^{-}_{j}+
O''^{L}_{ij}(\tilde{N}^{0}_{i})^{\dag}\bar{\sigma}^{\mu}\tilde{N}^{0}_{j}\right]Z_{\mu}\nonumber\\
&+&g\left[O^{L}_{ij}(\tilde{N}^{0}_{i})^{\dag}\bar{\sigma}^{\mu}\tilde{C}^{+}_{j}-O^{R}_{ij}(\tilde{N}^{0}_{i})\bar{\sigma}^{\mu}(\tilde{C}^{-}_{j})^{\dag}\right]W^{-}_{\mu}+c.c
\end{eqnarray}
with the definitions involved with the matrixes in \eqref{E3},
\begin{eqnarray}{\label{E4}}
O^{L}_{ij}&=& -\frac{1}{\sqrt{2}}N_{i4}V_{j2}^{*}+N_{i2}V_{j1}^{*}\nonumber\\
O^{R}_{ij}&=& -\frac{1}{\sqrt{2}}N_{i3}^{*}U_{j2}+N_{i2}^{*}U_{j1}\nonumber\\
O'^{L}_{ij}&=&-V_{i1}V^{*}_{j1} -\frac{1}{2}V_{i2}V_{j2}^{*}+\delta_{ij}s^{2}_{W}\\
O'^{R}_{ij}&=&-U_{i1}U^{*}_{j1} -\frac{1}{2}U_{i2}U_{j2}^{*}+\delta_{ij}s^{2}_{W}\nonumber\\
O''^{L}_{ij}&=&-O''^{R}_{ij}=\frac{1}{2}(N_{i4}N^{*}_{j4} -N_{i3}N_{j3}^{*})\nonumber
\end{eqnarray}
The Lagrangian \eqref{E3} gives rise to the relevant Feynman rules for the NC sector,
which are explicitly presented in the appendix A.
We can find that they agree with those shown in Fig. K.2.1  and Fig. K.2.2 in \cite{Martin08}.

\section{One-loop Contributions}

\begin{figure}[h!]
\centering
\begin{minipage}[b]{0.6\textwidth}
\centering
\includegraphics[width=4in]{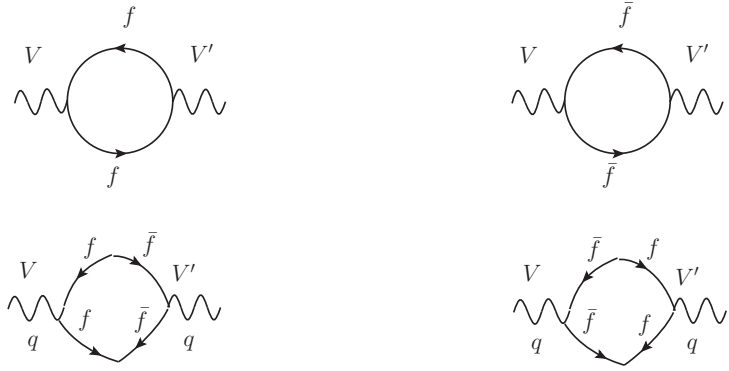}
\end{minipage}%
\caption{Graphs that contribute to the self-energy of SM  $\gamma$ and $Z$ bosons in the neutralino-chargino sector.}
\end{figure}

In terms of the Feynman rules for neutral vector fields $A_{\mu}$  and $Z_{\mu}$ coupled to the neutralinos and charginos  shown in the appendix A,
we see that the fermion pair in the fermion loop in fig.1 is either composed of $(\tilde{C}^{\pm}_{i},\tilde{C}^{\pm}_{j})$ or $(\tilde{N}^{0}_{i}, \tilde{N}^{0}_{j}) $.
In particular, only the neutral fermion pair in the fermion loop contributes to the self-energy of Z boson,
as in comparison with the self-energy of $\gamma$.
There are four Feynman diagrams for $i\Pi^{\gamma\gamma}$,
four Feynman diagrams for $i\Pi^{\gamma Z}$ and six diagrams for $i\Pi^{ZZ}$  in this sector.
Explicitly, these graphs give us \footnote{Note that there is an one-loop factor $16\pi^{2}$ multiplied by the $\Pi^{VV'}$ ignored throughout this section.},
 \begin{eqnarray}{\label{E5}}
\Pi^{\gamma\gamma}(q^{2})=4e^{2}[2A(q^{2}; m^{2}_{\tilde{C}^{+}_{i}},m^{2}_{\tilde{C}^{+}_{j}} ) -a(m^{2}_{\tilde{C}^{+}_{i}})+2q^{2}b_{0}(q^{2}; m^{2}_{\tilde{C}^{+}_{i}},m^{2}_{\tilde{C}^{+}_{j}})]
\end{eqnarray}
\begin{eqnarray}{\label{E6}}
\Pi^{\gamma Z}(q^{2})=\frac{eg}{c}(O'^{L}_{ii}+O'^{R}_{ii})
\left[-4A(q^{2}; m^{2}_{\tilde{C}^{+}_{i}},m^{2}_{\tilde{C}^{+}_{i}} )+2a(m^{2}_{\tilde{C}^{+}_{i}})-q^{2}b_{0}(q^{2}; m^{2}_{\tilde{C}^{+}_{i}},m^{2}_{\tilde{C}^{+}_{i}})\right]
\end{eqnarray}
and
\begin{eqnarray}{\label{E7}}
\Pi^{ZZ}(q^{2})&=&\frac{g^{2}}{c^{2}}\left\{\left(O'^{L}_{ij}O'^{L}_{ji}+O'^{R}_{ij}O'^{R}_{ji}\right)\left[4A(q^{2}; m^{2}_{\tilde{C}^{+}_{i}},m^{2}_{\tilde{C}^{+}_{j}} )-a(m^{2}_{\tilde{C}^{+}_{i}})-a(m^{2}_{\tilde{C}^{+}_{j}})\right.\right.\nonumber\\
&+&\left.\left(q^2-m^{2}_{\tilde{C}^{+}_{i}}-m^{2}_{\tilde{C}^{+}_{j}}\right)b_{0}(q^{2}; m^{2}_{\tilde{C}^{+}_{i}},m^{2}_{\tilde{C}^{+}_{j}})\right]\nonumber\\
 &+&\left.O''^{L}_{ij}O''^{L}_{ji} \left[4A(q^{2}; m^{2}_{\tilde{N}^{0}_{i}},m^{2}_{\tilde{N}^{0}_{j}} )-a(m^{2}_{\tilde{N}^{0}_{i}})-a(m^{2}_{\tilde{N}^{0}_{j}})\right.\right.\nonumber\\
 &+&\left.\left.\left(q^2-m^{2}_{\tilde{N}^{0}_{i}}-m^{2}_{\tilde{N}^{0}_{j}}\right)b_{0}(q^{2}; m^{2}_{\tilde{N}^{0}_{i}},m^{2}_{\tilde{N}^{0}_{j}})\right]\right\}\nonumber\\
&+&\frac{2g^{2}}{c^{2}}\left[2O'^{L}_{ij}O'^{R}_{ji}m_{\tilde{C}^{+}_{i}}m_{\tilde{C}^{+}_{j}}b_{0}(q^{2}; m^{2}_{\tilde{C}^{+}_{i}},m^{2}_{\tilde{C}^{+}_{j}})-
O''^{L}_{ij}O''^{L}_{ji}m_{\tilde{N}^{0}_{i}}m_{\tilde{N}^{0}_{j}}b_{0}(q^{2}; m^{2}_{\tilde{N}^{0}_{i}},m^{2}_{\tilde{N}^{0}_{j}})\right]\nonumber\\
\end{eqnarray}
For the definition of functions $A(q^{2}; x, y)$, $a(x)$ and $b_{0}(q^{2}; x, y)$,  we refer the reader to see appendix B in \cite{YZK}.

According to the property that W bosons only couple to $\chi^{0}_{i}$ and $\chi^{\pm}_{i}$,
not their anti-fermions, there is only one type of Feynman diagram as shown in fig. 2,
with $(\tilde{C}^{\pm}_{i}, \tilde{N}^{0}_{j})$ in the fermionic loop.
There are total four graphs needed to be evaluated in this situation.
\begin{figure}[h!]
\centering
\begin{minipage}[b]{0.6\textwidth}
\centering
\includegraphics[width=2in]{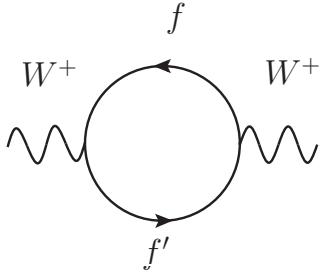}
\end{minipage}%
\caption{Graph that contributes to the self-energy of  $W$ boson the neutralino-chargino sector. }
\end{figure}
Using the Feynman rules shown in the appendix A yields
\begin{eqnarray}{\label{E8}}
\Pi^{WW}(q^{2})&=&g^{2}
\left\{\left[(O^{L}_{ij})^{*}O^{L}_{ij}+(O^{R}_{ij})^{*}O^{R}_{ij}\right]\left[4A(q^{2}; m^{2}_{\tilde{N}^{0}_{i}},m^{2}_{\tilde{C}^{+}_{j}} )-a(m^{2}_{\tilde{N}^{0}_{i}})-a(m^{2}_{\tilde{C}^{+}_{j}})\right.\right.\nonumber\\
&+&\left.\left.\left(q^2-m^{2}_{\tilde{N}^{0}_{i}}-m^{2}_{\tilde{C}^{+}_{j}}\right)b_{0}(q^{2}; m^{2}_{\tilde{N}^{0}_{i}},m^{2}_{\tilde{C}^{+}_{j}})\right]\right\}\\
&+&2g^{2}\left[(O^{L}_{ij})^{*}O^{R}_{ij}m_{\tilde{N}^{0}_{i}}m_{\tilde{C}^{+}_{j}}b_{0}(q^{2}; m^{2}_{\tilde{N}^{0}_{i}},m^{2}_{\tilde{C}^{+}_{j}})+
(O^{R}_{ij})^{*}O^{L}_{ij}m_{\tilde{N}^{0}_{i}}m_{\tilde{C}^{+}_{j}}b_{0}(q^{2}; m^{2}_{\tilde{N}^{0}_{i}},m^{2}_{\tilde{C}^{+}_{j}})\right]\nonumber
\end{eqnarray}

\section{A Preliminary Check via Decoupling Limit}
Now we proceed to perform a fast check on the results presented in the previous section.
Since the matrixes $O$'s appearing in \eqref{E5} to \eqref{E8} used to diagonalize the mass matrixes of neutralino $\mathcal{M}_{\tilde
{N}}$ and $\mathcal{M}_{\tilde{C}}$ are the main sources for the complication,
we take the large superpartner mass, i.e, the decoupling limit from the SM, to simplify these matrixes.

From the decoupling limit in which $M_{1,2}>> v_{\mu,d}$ and $\mu>> v_{\mu,d}$ ,
we obtain \cite{Group2A},
\begin{eqnarray}{\label{E9}}
O^{L}=O^{R}=\left(\begin{array}{cccc}
                   0 & 0 \\
                   1 & 0 \\
                   0 & -\frac{i}{2}\\
                    0 & \frac{1}{2}
                  \end{array}\right),~~~~~
O'^{L}=O'^{R}=\left(\begin{array}{cc}
                   s^{2}-1& 0 \\
                   0 & s^{2}-\frac{1}{2}
                  \end{array}\right)
\end{eqnarray}
and
\begin{eqnarray}{\label{E10}}
O''^{L}=O''^{R}=\left(\begin{array}{cccc}
                   0 & 0 &0&0 \\
                   0 & 0 &0&0\\
                   0&0& 0 & -\frac{i}{2}\\
                    0 &0 & \frac{i}{2}&0
                  \end{array}\right)
\end{eqnarray}
Substituting \eqref{E5} to \eqref{E8}, \eqref{E9} and \eqref{E10} into the difference
between the two definitions of weak mixing angle squared $s^{2}$ and $s^{2}_{*}$ results in,
\begin{eqnarray}{\label{E11}}
s^{2}-s^{2}_{*}&\equiv&-\frac{c^{2}}{m_{W}^{2}}\left[\Pi^{WW}(m^{2}_{W})-c^{2}\Pi^{ZZ}(m^{2}_{Z})\right]
+\frac{sc^{3}}{m_{W}^{2}}\Pi^{\gamma Z}(m^{2}_{Z})\nonumber\\
&=&\left[2(m_{\tilde{N}^{0}_{2}}-m_{\tilde{C}^{+}_{1}})^{2}
+\frac{1}{2}(m_{\tilde{N}^{0}_{3}}-m_{\tilde{C}^{+}_{2}})^{2}
\right.\nonumber\\
&+&\left.\frac{1}{2}(m_{\tilde{N}^{0}_{4}}-m_{\tilde{C}^{+}_{2}})^{2}
-\frac{1}{2}(m_{\tilde{N}^{0}_{3}}-m_{\tilde{N}^{0}_{4}})^{2}\right]\eta +\cdots
\end{eqnarray}
In \eqref{E11}, we have ignored the finite terms.
By using the fact that $m_{\tilde{C}^{+}_{2}}=m_{\tilde{N}^{0}_{3}}=m_{\tilde{N}^{0}_{4}}=\mu$ and
$m_{\tilde{C}^{+}_{1}}=m_{\tilde{N}^{0}_{2}}=M_{2}$ ($M_2$ denotes the second gaugino mass) under the decoupling limit,
we arrive at the conclusion that the divergent parts in \eqref{E11} cancel exactly.

How about the finite property of \eqref{E11} on general grounds?
The answer to this question is unclear due to the complication that the matrixes as coefficients in the
results \eqref{E5} to \eqref{E8} are tied to the gaugino masses $M_{1,2}$ and $\mu$ term.
Without the information about these soft masses, one can not determine them generally.

\section{Estimates of $S$, $T$, $U$ Parameters}
In this section, we derive the NC sector's contribution to parameters  $S$, $T$ and $U$ \cite{Peskin1, Peskin2},
which measure the oblique corrections to precise electroweak observables.
The dependence of  $S$, $T$ and $U$ parameters on  $\Pi^{IJ}(p^{2})$ is given by \cite{Peskin1, Peskin2},
 \begin{eqnarray}{\label{E12}}
S&\equiv&-\frac{16\pi}{e^{2}}sc\left[sc\Pi^{\gamma\gamma '}(0)-sc\Pi^{ZZ '}(0)+(c^{2}-s^{2})\Pi^{\gamma Z '}(0)\right]\nonumber\\
T&\equiv&\frac{4\pi}{e^{2}}\left[\frac{\Pi^{WW}(0)}{m^{2}_{W}}-\frac{\Pi^{ZZ}(0)}{m^{2}_{Z}}-\frac{2s}{c}\frac{\Pi^{\gamma Z}(0)}{m^{2}_{Z}}\right]\\
U&\equiv&\frac{16\pi s^{2}}{e^{2}}\left[\Pi^{WW '}(0)-c^{2}\Pi^{ZZ '}(0)-s^{2}\Pi^{\gamma\gamma '}(0)-2c
s\Pi^{\gamma Z '}(0)\right]\nonumber
\end{eqnarray}
with  $\Pi^{IJ'}(0)=d^{2}\Pi^{IJ}/dp^{2}\mid_{p^{2}=0}$,
here $\Pi^{IJ}$ represents the part with metric as the coefficient in $\Pi_{\mu\nu}^{IJ}=g_{\mu\nu}\Pi^{IJ}+\cdots$.

Substitute \eqref{E5} to \eqref{E8} into \eqref{E12} gives rise to
\footnote{Note that the summation over index $i$ and $j$ is performed throughout \eqref{E13} to \eqref{E15}.},
 \begin{eqnarray}{\label{E13}}
\pi S_{NC}&=&-s^{2}c^{2}\left[8A'(0; m^{2}_{\tilde{C}_{i}^{+}}, m^{2}_{\tilde{C}_{i}^{+}})+2b_{0}(0; m^{2}_{\tilde{C}_{i}^{+}}, m^{2}_{\tilde{C}_{i}^{+}})\right]\nonumber\\
&+&O'^{L(R)}_{ij}O'^{L(R)}_{ji}\left[4A'(0; m^{2}_{\tilde{C}_{i}^{+}}, m^{2}_{\tilde{C}_{j}^{+}})+b_{0}(0; m^{2}_{\tilde{C}_{i}^{+}}, m^{2}_{\tilde{C}_{j}^{+}})-\left(m^{2}_{\tilde{C}_{i}^{+}}+m^{2}_{\tilde{C}_{j}^{+}}\right)b'_{0}(0; m^{2}_{\tilde{C}_{i}^{+}}, m^{2}_{\tilde{C}_{j}^{+}})\right]\nonumber\\
&+&4 O'^{L}_{ij}O'^{R}_{ji}m_{\tilde{C}_{i}^{+}}m_{\tilde{C}_{j}^{+}}b'_{0}(0; m^{2}_{\tilde{C}_{i}^{+}}, m^{2}_{\tilde{C}_{j}^{+}})
-2O''^{L}_{ij}O''^{L}_{ji}m_{\tilde{N}_{i}^{0}}m_{\tilde{N}_{j}^{0}}b'_{0}(0; m^{2}_{\tilde{N}_{i}^{0}}, m^{2}_{\tilde{N}_{j}^{0}})\nonumber\\
&+&O''^{L}_{ij}O''^{L}_{ji}\left[4A'(0; m^{2}_{\tilde{N}_{i}^{0}}, m^{2}_{\tilde{N}_{j}^{0}})
+b_{0}(0; m^{2}_{\tilde{N}_{i}^{0}}, m^{2}_{\tilde{N}_{j}^{0}})-\left(m^{2}_{\tilde{C}_{i}^{+}}+m^{2}_{\tilde{C}_{j}^{+}}\right)b'_{0}(0; m^{2}_{\tilde{C}_{i}^{+}}, m^{2}_{\tilde{C}_{j}^{+}})\right]\nonumber\\
&+&(c^{2}-s^{2})\left(O'^{L}_{ii}+O'^{R}_{ii}\right)\left(4A'(0; m^{2}_{\tilde{C}_{i}^{+}}, m^{2}_{\tilde{C}_{i}^{+}})+b_{0}(0; m^{2}_{\tilde{C}_{i}^{+}}, m^{2}_{\tilde{C}_{i}^{+}})\right)
\end{eqnarray}
 \begin{eqnarray}{\label{E14}}
\pi T_{NC}&=&\frac{1}{4s^{2}m^{2}_{W}}\left\{\left[(O^{L}_{ij})^{*}O^{L}_{ij}+(O^{R}_{ij})^{*}O^{R}_{ij}\right]\left[4A(0; m^{2}_{\tilde{N}_{i}^{0}}, m^{2}_{\tilde{C}_{j}^{+}})-a(m^{2}_{\tilde{N}_{i}^{0}})-a(m^{2}_{\tilde{C}_{j}^{+}})\right.\right.\nonumber\\
&-&\left.\left.\left(m^{2}_{\tilde{N}_{i}^{0}}+m^{2}_{\tilde{C}_{j}^{+}}\right)b_{0}(0; m^{2}_{\tilde{N}_{i}^{0}}, m^{2}_{\tilde{C}_{j}^{+}})\right]+2\left((O^{L}_{ij})^{*}O^{R}_{ij}+(O^{R}_{ij})^{*}O^{L}_{ij}\right)m_{\tilde{N}_{i}^{0}}m_{\tilde{C}_{j}^{+}}b_{0}(0; m^{2}_{\tilde{N}_{i}^{0}}, m^{2}_{\tilde{C}_{j}^{+}})\right.\nonumber\\
&-&\left.\left(O'^{L}_{ij}O'^{L}_{ji}+O'^{R}_{ij}O'^{R}_{ji}\right)\left[4A(0; m^{2}_{\tilde{C}_{i}^{+}}, m^{2}_{\tilde{C}_{j}^{+}})-a(m^{2}_{\tilde{C}_{i}^{+}})-a(m^{2}_{\tilde{C}_{j}^{+}})\right.\right.\nonumber\\
&-&\left.\left.\left(m^{2}_{\tilde{C}_{i}^{+}}+m^{2}_{\tilde{C}_{j}^{+}}\right)b_{0}(0; m^{2}_{\tilde{C}_{i}^{+}}, m^{2}_{\tilde{C}_{j}^{+}})\right]-4O'^{L}_{ij}O'^{R}_{ji}m_{\tilde{N}_{i}^{0}}m_{\tilde{C}_{j}^{+}}b_{0}(0; m^{2}_{\tilde{N}_{i}^{0}}, m^{2}_{\tilde{C}_{j}^{+}})\right.\nonumber\\
&-&\left.O''^{L}_{ij}O''^{L}_{ji}
\left[4A(0; m^{2}_{\tilde{N}_{i}^{0}}, m^{2}_{\tilde{N}_{j}^{0}})-a(m^{2}_{\tilde{N}_{i}^{0}})-a(m^{2}_{\tilde{N}_{0}^{+}})-\left(m^{2}_{\tilde{N}_{i}^{0}}+m^{2}_{\tilde{N}_{j}^{0}}\right)b_{0}(0; m^{2}_{\tilde{N}_{i}^{0}}, m^{2}_{\tilde{N}_{j}^{0}})\right]\right.\nonumber\\
&+&\left.2s^{2}\left(O'^{L}_{ii}+O'^{R}_{ii}\right)\left[4A(0; m^{2}_{\tilde{C}_{i}^{+}}, m^{2}_{\tilde{C}_{i}^{+}})-2a(m^{2}_{\tilde{C}_{i}^{+}})\right]+2O''^{L}_{ij}O''^{L}_{ji}m_{\tilde{N}_{i}^{0}}m_{\tilde{N}_{j}^{0}}b_{0}(0; m^{2}_{\tilde{N}_{i}^{0}}, m^{2}_{\tilde{N}_{j}^{0}})\right\}\nonumber\\
\end{eqnarray}
and
\begin{eqnarray}{\label{E15}}
\pi U_{NC}&=&\left[(O^{L}_{ij})^{*}O^{L}_{ij}+(O^{R}_{ij})^{*}O^{R}_{ij}\right]\left[4A'(0; m^{2}_{\tilde{N}_{i}^{0}}, m^{2}_{\tilde{C}_{j}^{+}})-\left(m^{2}_{\tilde{N}_{i}^{0}}+m^{2}_{\tilde{C}_{j}^{+}}\right)b'_{0}(0; m^{2}_{\tilde{N}_{i}^{0}}, m^{2}_{\tilde{C}_{j}^{+}})\right.\nonumber\\
&+&\left.b_{0}(0; m^{2}_{\tilde{N}_{i}^{0}}, m^{2}_{\tilde{C}_{j}^{+}})\right]+2\left[(O^{L}_{ij})^{*}O^{R}_{ij}+(O^{R}_{ij})^{*}O^{L}_{ij}\right]m_{\tilde{N}_{i}^{0}}m_{\tilde{C}_{j}^{+}}b_{0}(0; m^{2}_{\tilde{N}_{i}^{0}}, m^{2}_{\tilde{C}_{j}^{+}})\nonumber\\
&-&O''^{L}_{ij}O''^{L}_{ji}\left[4A'(0; m^{2}_{\tilde{N}_{i}^{0}}, m^{2}_{\tilde{N}_{j}^{0}})+b_{0}(0; m^{2}_{\tilde{N}_{i}^{0}}, m^{2}_{\tilde{C}_{j}^{+}})-\left(m^{2}_{\tilde{N}_{i}^{0}}+m^{2}_{\tilde{N}_{j}^{0}}\right)b'_{0}(0; m^{2}_{\tilde{N}_{i}^{0}}, m^{2}_{\tilde{N}_{j}^{0}})\right]\nonumber\\
&-&\left[O'^{L}_{ij}O'^{L}_{ji}+O'^{R}_{ij}O'^{R}_{ji}\right]\left[4A'(0; m^{2}_{\tilde{c}_{i}^{+}}, m^{2}_{\tilde{C}_{j}^{+}})-\left(m^{2}_{\tilde{C}_{+}^{0}}+m^{2}_{\tilde{C}_{j}^{+}}\right)b'_{0}(0; m^{2}_{\tilde{C}_{i}^{+}}, m^{2}_{\tilde{C}_{j}^{+}})\right.\nonumber\\
&+&\left.b_{0}(0; m^{2}_{\tilde{C}_{i}^{+}},m^{2}_{\tilde{C}_{j}^{+}})\right]+2s^{2}\left(O'^{L}_{ii}+O'^{R}_{ii}-s^{2}\right)\left[4A'(0; m^{2}_{\tilde{C}_{i}^{+}}, m^{2}_{\tilde{C}_{i}^{+}})+b_{0}(0; m^{2}_{\tilde{C}_{i}^{+}}, m^{2}_{\tilde{C}_{j}^{+}})\right]\nonumber\\
&-&4O'^{L}_{ij}O'^{R}_{ji}m_{\tilde{C}_{i}^{+}}m_{\tilde{C}_{j}^{+}}b'_{0}(0; m^{2}_{\tilde{C}_{i}^{+}}, m^{2}_{\tilde{C}_{j}^{+}})+2O''^{L}_{ij}O''^{L}_{ji}m_{\tilde{N}_{i}^{0}}m_{\tilde{N}_{j}^{0}}b'_{0}(0; m^{2}_{\tilde{N}_{i}^{0}}, m^{2}_{\tilde{N}_{j}^{0}})
\end{eqnarray}
From \eqref{E13} to \eqref{E15} one can see that the finite property of parameters $S$, $T$ and $U$ is obvious.
If we assume that the SUSY mass splitting between any two mass parameters in the set of $m_{\tilde{N}_{i}}$ and $m_{\tilde{C}_{j}}$ is small compared with themselves, the results in \eqref{E13} to \eqref{E15} can be further simplified  as,
\begin{eqnarray}{\label{E16}}
\pi
S_{NC}&\simeq&\frac{1}{3}\left[\ln\frac{ m^{2}_{\tilde{N}_{3}^{0}}}{ m^{2}_{\tilde{C}_{2}^{+}}}
+ \frac{m^{2}_{\tilde{N}_{4}^{0}}-m^{2}_{\tilde{N}_{3}^{0}}}{2m^{2}_{\tilde{N}_{4}^{0}}}\right]
\end{eqnarray}

\begin{eqnarray}{\label{E17}}
\pi T_{NC}&\simeq&\frac{1}{16s^{2}m^{2}_{W}}\left[-4\left(\frac{m^{2}_{\tilde{C}_{1}^{+}}-m^{2}_{\tilde{N}_{2}^{0}}}{m^{2}_{\tilde{C}_{1}^{+}}}\right)^2m^{2}_{\tilde{N}_{2}^{0}}\left(\frac{1}{2}\ln\frac{ m^{2}_{\tilde{N}_{2}^{0}}}{\Lambda^2}+\frac{1}{3}\right)\right.\nonumber\\ &-&\left.\left(\frac{m^{2}_{\tilde{C}_{2}^{+}}-m^{2}_{\tilde{N}_{3}^{0}}}{m^{2}_{\tilde{C}_{2}^{+}}}\right)^2m^{2}_{\tilde{N}_{3}^{0}}\left(\frac{1}{2}\ln\frac{ m^{2}_{\tilde{N}_{3}^{0}}}{\Lambda^2}+\frac{1}{3}\right)\right.\\
&-&\left.\left(\frac{m^{2}_{\tilde{C}_{2}^{+}}-m^{2}_{\tilde{N}_{4}^{0}}}{m^{2}_{\tilde{C}_{2}^{+}}}\right)^2m^{2}_{\tilde{N}_{4}^{0}}\left(\frac{1}{2}\ln\frac{ m^{2}_{\tilde{N}_{4}^{0}}}{\Lambda^2}+\frac{1}{3}\right)
+\left(\frac{m^{2}_{\tilde{N}_{4}^{0}}-m^{2}_{\tilde{N}_{3}^{0}}}{m^{2}_{\tilde{N}_{4}^{0}}}\right)^2m^{2}_{\tilde{N}_{3}^{0}}\left(\frac{1}{2}\ln\frac{ m^{2}_{\tilde{N}_{3}^{0}}}{\Lambda^2}+\frac{1}{3}\right)\right]\nonumber
\end{eqnarray}
together with
\begin{eqnarray}{\label{E18}}
\pi U_{NC}&\simeq&\frac{1}{6}\left[8\ln\frac{m^{2}_{\tilde{N}_{2}^{0}}}{m^{2}_{\tilde{C}_{1}^{+}}}+2\ln \frac{m^{2}_{\tilde{N}_{4}^{0}}}{m^{2}_{\tilde{C}_{2}^{+}}}
+4\frac{m^{2}_{\tilde{C}_{1}^{+}}-m^{2}_{\tilde{N}_{2}^{0}}}{2m^{2}_{\tilde{C}_{1}^{+}}}\right.\nonumber\\
&+&\left.\frac{m^{2}_{\tilde{C}_{2}^{+}}-m^{2}_{\tilde{N}_{4}^{0}}}{m^{2}_{\tilde{C}_{2}^{+}}}
+\frac{m^{2}_{\tilde{C}_{2}^{+}}-m^{2}_{\tilde{N}_{3}^{0}}}{m^{2}_{\tilde{C}_{2}^{+}}}-\frac{m^{2}_{\tilde{N}_{4}^{0}}-m^{2}_{\tilde{N}_{3}^{0}}}{m^{2}_{\tilde{N}_{4}^{+}}}
\right]\nonumber\\
\end{eqnarray}
by using the approximations
\begin{eqnarray}{\label{E19}}
A(0;m_{1}^{2},m_{2}^{2})&\simeq&-\frac{1}{2}m_{1}^{2}+\frac{1}{2}m_{1}^{2}(1+\frac{t}{2})\ln\frac{m_{1}^{2}}{\Lambda^2}+\frac{1}{12}m_{1}^{2}t^2\left(1+3\ln\frac{m_{1}^{2}}{\Lambda^2}\right)\nonumber\\
b_{0}(0; m_{1}^{2}, m_{2}^{2})&\simeq&\ln\frac{ m_{1}^{2}}{\Lambda^2}+\frac{t}{2}\nonumber\\
A'(0; m_{1}^{2}, m_{2}^{2})&\simeq&-\frac{1}{12}\ln\frac{ m_{1}^{2}}{\Lambda^2}-\frac{t}{24} \\
b'_{0}(0; m_{1}^{2}, m_{2}^{2})&\simeq&\frac{1}{m_{1}^{2}}\left(-\frac{1}{6}+\frac{t}{12} \right) \nonumber
\end{eqnarray}
for $\mid m_{1}^{2}-m_{2}^{2}\mid<< m_{1}^{2}$ , $\mid m_{1}^{2}-m_{2}^{2}\mid<< m_{2}^{2}$  and  $t=(m_{2}^{2}-m_{1}^{2})/m_{2}^{2}$.
$\Lambda$ is the usual scale mass of dimensional regularization,
as introduced in the integral functionals.\\

\section{Discussions and Conclusions}
This paper is devoted to revisit calculating oblique corrections in the context of MSSM at one-loop.
The theoretic motivation for this effort is due to discrepancies among the results presented in various earlier works.
In contrast with those in the literature  \cite{SUSY, Group2A, Group3},
we take the two-component spinor formalism to perform the calculation in the neuralino-chargino sector.
The final results of one-loop $\Pi^{VV'}$ are examined in the large SUSY mass limit.
which are further verified by the finite $S$, $T$ and $U$ parameters induced by $\Pi^{VV'}$ terms.

In comparison with those in \cite{SUSY} (See also \cite{Group3}),
we find they \em{exactly}\em~ coincide with our $\Pi^{VV'}$
despite the factors involved in the matrixes in \eqref{E4} as the coupling coefficients,
by using the relations between the $B_{3}$ and $B_{4}$ functionals in \cite{SUSY} and $a$, $A$ and $b_{0}$ functionals in this note.
In comparison with the results of $S$, $T$, $U$ parameters in \cite{Group2A} (where weakly broken $SU(2)$ symmetries assumed),
we find they are not agree with \eqref{E16}-\eqref{E18}.
To simplify $T$ in \eqref{E17}, one can redefine the mass scale
$\Lambda\rightarrow \tilde{\Lambda}$ to absorb the factor $1/3$.
Consequently, $T$ is a logarithm function in structure, as the same with that in \cite{Group2A}.
But the coefficients of these logarithm terms do not agree.

In summary, the correct estimate of one-loop oblique corrections
in the context of MSSM is obtained in terms of two-component spinor formalism.
By incorporating with the bosonic contributions obtained in the previous work \cite{YZK},
one can apply the oblique correction as a portal to examine the MSSM in light of recent LHC data on SUSY .\\

~~~~~~~~~~~~~~~~~~~~~~~~~~~~~~~~~~~~
~~~$\bf{Acknowledgement}$\\
SZ thanks Ken-ichi Hikasa and Jin Min Yang for communication.
The work of YY is supported by the Fundamental Research Funds for the Central Universities under Grant No.CDJXS1102209,
that of SZ is supported in part by the Doctoral Fund of Ministry of Education of China (No. 20110191120045).

\newpage
\appendix
\section{Relevant Feynman Rules in the NC sector}
\begin{figure}[h!]
\centering
\begin{minipage}[b]{0.8\textwidth}
\centering
\includegraphics[width=5in]{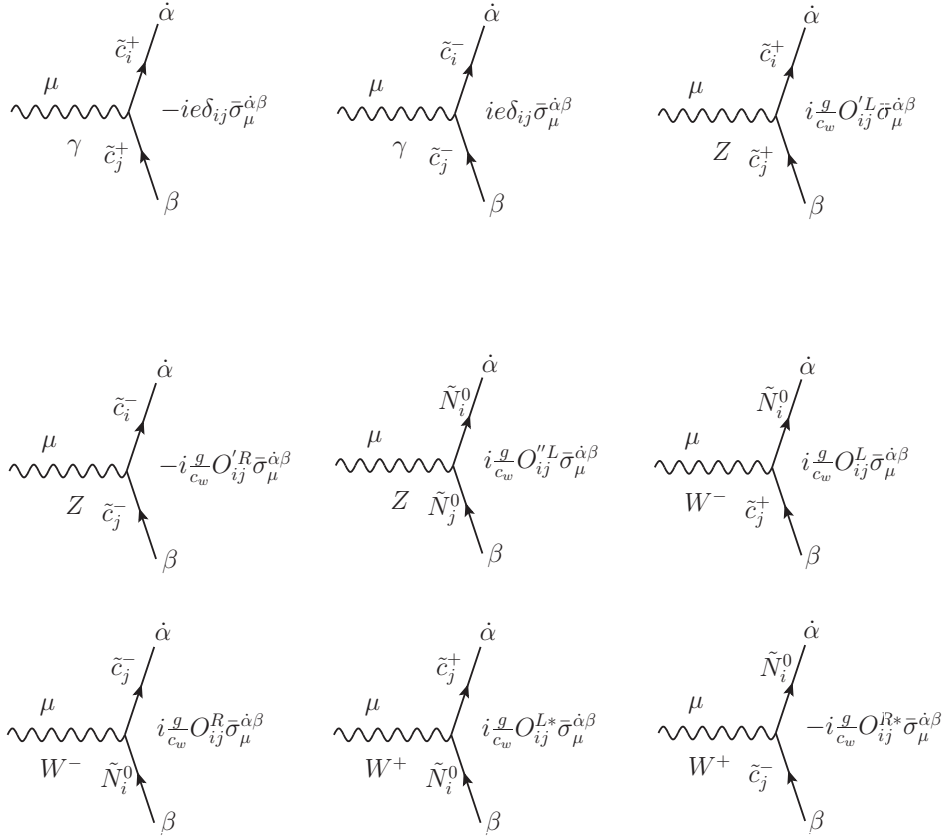}
\end{minipage}%
\caption{Relevant Feynman rules in the NC sector in two-component formalism. }
\end{figure}

\end{document}